\documentclass[runningheads]{llncs}

\usepackage{newtxtext}
\usepackage{newtxmath} 
\usepackage[scaled=0.88,semibold,extracondensed]{noto-mono} 
\usepackage[scaled=0.88,semibold,semicondensed]{noto-sans} 
\usepackage[T1]{fontenc}

\usepackage{xspace}
\usepackage[fleqn]{mathtools}
\usepackage{hyperref}
\usepackage{tabularx}
\usepackage{gensymb}

\RequirePackage{color}
\definecolor{DodgerUniformBlue}{rgb}{0.0,0.353,0.612}
\newcommand{\define}[1]{\emph{\textcolor{DodgerUniformBlue}{#1}}}

\newcommand{\pbox}[2][c]{\begin{tabular}[#1]{@{}l@{}}#2\end{tabular}}

\RequirePackage{tabularx}%
\newcolumntype{L}{>{$}l<{$}}%
\newcolumntype{C}{>{$}c<{$}}%
\newcolumntype{R}{>{$}r<{$}}%
\newcolumntype{M}{@{}p{\mathindent}@{}}%
\newcolumntype{t}{>{\mbox\bgroup}l<{\egroup}}%
\newcolumntype{e}{r@{\;}l}%
\newcolumntype{E}{>{$}r<{$}@{$\;$}>{$}l<{$}}%

\usepackage{listings}
\let\LIN=\lstinline

\newcommand{\Time}{\mathbb{T}}

\newcommand{\ALCIF}{\ensuremath{\cal ALCIF}\xspace}
\newcommand{\ALC}{\ensuremath{\cal ALC}\xspace}

\renewcommand{\:}[1][{}]{\,{:#1}\,}

\def\funfont{\upshape\sffamily\mdseries}
\def\fun#1{\text{\funfont #1}}

\def\IsAAt{\fun{IsAAt}}
\def\HasAAt{\fun{HasAAt}}

\newcommand{\colonminus}{\mathrel{\fun{:--}}}

\def\kwfont{\upshape\sffamily\bfseries}
\newcommand{\kw}[1]{\text{\kwfont #1}}

\DeclareMathOperator{\NOT}{\kw{not}}
\DeclareMathOperator{\FAIL}{\kw{fail}}

\DeclareMathOperator{\LET}{\kw{let}}
\DeclareMathOperator{\CHOOSE}{\kw{choose}}
\DeclareMathOperator{\COLLECT}{\kw{collect}}
\DeclareMathOperator{\NEG}{\kw{neg}}

\DeclareMathOperator{\STH}{\kw{sth}}

\DeclareMathOperator{\var}{\mathit{var}}

\DeclareMathOperator{\fvar}{\mathit{fvar}}

\newcommand{\timevar}{\mathit{time}}
\newcommand{\timeterm}{\mathit{tt}}
\newcommand{\TT}{\timeterm}
\newcommand{\TV}{\timevar}
\newcommand{\AT}{\mathbin{@}}

\DeclareMathOperator{\dom}{\mathit{dom}}

\begin{document}
\title{Combining Event Calculus and Description Logic Reasoning via Logic Programming}

\author{Peter Baumgartner}
\authorrunning{P. Baumgartner}

\institute{Data61/CSIRO and The Australian National University, Canberra, Australia
\href{mailto:Peter.Baumgartner@data61.csiro.au}{\email{Peter.Baumgartner@data61.csiro.au}}}
\maketitle              
\begin{abstract}
  The paper introduces a knowledge representation language that combines the event
  calculus with description logic in a logic programming framework. The purpose is to
  provide the user with an expressive language for modelling and analysing systems that
  evolve over time.
  The approach is exemplified with the logic programming language as implemented in the
  Fusemate system. The paper extends Fusemate's rule language with a weakly DL-safe
  interface to the description logic \ALCIF and adapts the event calculus to this extended language.
  This way, time-stamped ABoxes can be manipulated as fluents in the event
  calculus. All that is done in the frame of Fusemate's concept of stratification by time.
  The paper provides conditions for soundness and completeness where appropriate.
  Using an elaborated example it demonstrates the interplay of the event calculus,
  description logic and
  logic programming rules for computing possible models as plausible explanations of the current state
  of the modelled system.\\
\textbf{This is a corrected version of the published paper. It adds a missing case in the definition of the semantics
  of body literals (Section~\ref{sec:possible-models}), and it fixes a flaw in the
  definition of possible models (See Note~\ref{note:flaw}).}
  \end{abstract}

\section{Introduction}
\label{sec:introduction}
This paper presents an expressive logical language for modelling systems that evolve over
time. The language is intended for model computation: given a history of events until
``now'', what are the system states at these times, in particularly ``now'', expressed as
logical models. This is a useful reasoning service in application areas with only
partially observed events or incomplete domain knowledge.  By making informed guesses and
including its consequences, the models are meant to provide plausible explanations for
helping understand the current issues, if any, as a basis for further decision making.

For example, transport companies usually do not keep detailed records of what goods went
on what vehicle for a transport on a particular day. Speculating the whereabouts of a
missing item can be informed by taking known locations of other goods of the same
batch on that day into account; problems observed with goods on delivery site, e.g., low quality of
fresh goods, may or may not be related to the transport conditions, and playing through
different scenarios may lead to plausible explanations while eliminating others (truck cooling problems?
tampering?).

There are numerous approaches for modelling and analysing systems that evolve over time. They
are often subsumed under the terms of stream processing, complex event recognition, and
situational awareness, temporal verification among others,
see~\cite{DBLP:journals/ker/ArtikisSPP12,DBLP:journals/ai/BeckDE18,Baader:etal:SAIL:TABLEAUX:2009,DBLP:conf/ausai/BaaderBL15,Baader:Ghilardi:LTL-over-Description-Logic-Axioms:ACM:2008}
for some logic-based methods.
Symbolic event recognition, for instance, accepts as input a stream of time-stamped
low-level events and identifies high-level events — collections of events that satisfy
some pattern~\cite{DBLP:journals/ker/ArtikisSPP12}. See~\cite{10.1145/3328905.3329504} for
a recent sophisticated event calculus.
Other approaches utilize description logics in a temporalized setting of ontology-based
data access (OBDA)~\cite{OBDA:Poggi:etal:2008}.
For instance, \cite{10.1007/978-3-319-11206-0_18} describes a method for streaming data
into a sequence  of ABoxes, which can be queried in an SQL-like language with respect
to a given ontology. 

  The knowledge representation language put forward in this paper combines Kowalski's
  event calculus (EC) with description logics (DL) in a logic programming framework.  The
  rationale is, DLs have a long history of developments for representing structured domain
  knowledge and for offering reliable (decidable) reasoning services.  The EC provides a
  structured way of representing actions and their effects, represented as fluents that
  may change their truth value over time.  For the intended model computation applications
  mentioned above, the EC makes it easy to take snapshots of the fluents at any time.  The
  full system state at a chosen time then is derived from the fluent snapshot and DL
  reasoning. The logic programming rules orchestrate their integration and serve other
  purposes, such as diagnosis.

 This paper uses the Fusemate logic programming language and
system~\cite{Baumgartner:PossibleModelsSpringer:IJCAR:2020,Baumgartner:Fusemate:SystemDescription:CADE:2021}. Fusemate 
computes possible models of stratified disjunctive logic
programs~\cite{Sakama:PossibleModelSemanticsDisjunctiveDatabases:DOOD:89,Sakama:Inoue:PossibleModels:JAR:94}, see
Section~\ref{sec:stratified-logic-programs} for details.
Fusemate was introduced in~\cite{Baumgartner:PossibleModelsSpringer:IJCAR:2020} with the
same motivation as here. In~\cite{Baumgartner:Fusemate:SystemDescription:CADE:2021} it was
extended with novel language operators improved with a weaker form of stratification.
Their usefulness in combination was demonstrated by application to description logic
reasoning. In~\cite{Baumgartner:Fusemate:SystemDescription:CADE:2021}  it
was shown how to
transform an \ALCIF\footnote{\ALCIF is the well-known description logic \ALC extended with
  inverse roles and functional roles. See~\cite{Baader:etal:introduction-description-logics:2017} for
  background on description logics.} knowledge base into a set of
Fusemate rules and facts that is satisfiable if and only if the knowledge base is
\ALCIF-satisfiable. All of that is used in this paper.

\paragraph{Paper contributions.}
This paper builds on the Fusemate developments summarized above and extends them in the following ways:
\begin{enumerate}
\item Integration of the description logic reasoner
  of~\cite{Baumgartner:Fusemate:SystemDescription:CADE:2021} as a subroutine callable from
  Fusemate rules. 
  Section~\ref{sec:dl-interface} details the semantics of the combination and 
  conditions for its soundness and completeness. This is an original contribution in its own
  right 
  which exploits advantages of a stratified setting.
\item A version of the event calculus~\cite{DBLP:journals/ngc/KowalskiS86}
  that fits Fusemate's model computation and notion of stratification.
  Details are in  Section~\ref{sec:event-calculus}.
\item Integrating DL and EC by means of rules, and utilizing rules for KR aspects not
  covered by either.  Details in particular in Section~\ref{sec:alltogether}
\item Providing an elaborated example  the integrated
  EC/DL/rules language. It is included in the Fusemate distribution which is available at \url{https://bitbucket.csiro.au/users/bau050/repos/fusemate/}.
\end{enumerate}
To the best of my knowledge, a combination of DL with EC has not been considered before.
Given the long history of applying DL reasoning (also) for time evolving systems, I find
this surprising.  From that perspective, a main contribution of this paper is to fill the
gap and to argue that the proposed combination makes sense.

There is work is on integrating DLs into 
the situation calculus (SitCalc) and similar methods~\cite{Integrating-Action-Calculi-and-Description-Logics:Drescher:Thielscher:KI:2007,Integrating-Description-Logics-and-Action-Formalisms:Baader:etal:AAAI:2005,DL-ramification-problem:BaaderLetal:LPAR:2010,Description-Logic-Knowledge-and-Action-Bases:Hariri:etal:JAIR:2013}.
SitCalc~\cite{DBLP:reference/fai/Lin08}  is a first-order logic formalism for specifying state transitions
in terms of pre- and post-conditions of actions. Its is
mostly used for planning and related applications that require reachability reasoning for state transitions.
Indeed, the
papers~\cite{Integrating-Description-Logics-and-Action-Formalisms:Baader:etal:AAAI:2005}
and~\cite{Integrating-Action-Calculi-and-Description-Logics:Drescher:Thielscher:KI:2007}
investigate reasoning tasks (executability and projection, ABox updates) that are relevant in that
context. Both approaches are restricted to acyclic TBoxes.
In~\cite{Description-Logic-Knowledge-and-Action-Bases:Hariri:etal:JAIR:2013}, 
actions are speciﬁed as sets 
of conditional effects, where conditions are based on epistemic queries over the knowledge base
(TBox and ABox), and effects are expressed in terms of new ABoxes. The paper investigates
veriﬁcation of temporal properties. As a difference to the EC, none of these approaches supports
a \emph{quantitative} notion of time.

\section{Stratified Logic Programs and Model Computation}
\label{sec:stratified-logic-programs}
This paper uses the extended ``Fusemate'' rule language introduced
in~\cite{Baumgartner:Fusemate:SystemDescription:CADE:2021} without the earlier belief 
revision operator introduced in~\cite{Baumgartner:PossibleModelsSpringer:IJCAR:2020}.
This section complements the earlier paper~\cite{Baumgartner:Fusemate:SystemDescription:CADE:2021}
with a rigorous definition of the semantics of the extended language. It also provides
soundness and completeness arguments, under certain conditions, wrt.\ abstract fixpoint
iteration and wrt.\ Fusemate's procedure more concretely.

Terms and atoms of a given first-order signature with ``free'' \define{ordinary} function and predicate symbols are defined as usual.
Let $\Time$ be a countably infinite discrete set of \define{time points} equipped with a
well-founded total strict ordering $<$ (strictly earlier), e.g., the natural numbers. Assume that the time
points, comparison operators $=$, $\le$ (earlier),  and a next time function  $+ 1$ are also
part of the signature and interpreted in the obvious way.
A \define{time term} is a (possibly non-ground) term  over the sub-signature $\Time \cup \{+1\}$. 
The signature may contain other ``built-in'' interpreted predicate and function symbols for predefined
types such as strings, arithmetic data types, sets, etc.  We only informally assume that
all terms are built in a well-sorted way, and that interpreted operators over ground terms can
be evaluated effectively to a value represented by a term.

Let $\var(e)$ denote the set of
variables occurring in a term or atom $e$. We say that $e$ is \define{ground} if $\var(e) = \emptyset$.  
We write $e\sigma$ for applying a substitution $\sigma$ to
$e$. The domain of $\sigma$ is denoted by $\dom(\sigma)$.
A substitution $\gamma$ is a \define{grounding substitution for a finite set of variables $X$} iff $\dom(\gamma) = X$.
In the following,  the letters $x,y,z$ stand for variables, $\timevar$ for a time term
variable, $s, t$ for terms, and $\timeterm$ for a time term, possibly
indexed. Lists of terms or other expressions are written as vectors, e.g., $\vec t$ is a list of terms
$t_1,\ldots,t_n$ for some $n >= 0$.
A \define{(Fusemate) rule} is an implication written in Prolog-like syntax as
    \begin{equation}
      \label{eq:rule}
      H \colonminus b_1,\ldots,b_k, \NOT \vec{b}_1, \ldots ,\NOT \vec{b}_n \enspace.
    \end{equation}
In~\eqref{eq:rule}, the rule \define{head} $H$ is either (a) \define{ordinary}, a disjunction $h_1 \vee \cdots \vee h_m$ of ordinary atoms, for
    some $m \ge 1$, or (b) the expression $\FAIL$.\footnote{This definition of head is
      actually simplified as Fusemate offers an additional head 
      operator  for belief revision,
      see\cite{Baumgartner:PossibleModelsSpringer:IJCAR:2020}. This is ignored here.} In
    case (a) the rule is
    \define{ordinary} and in case (b) it is a \define{fail rule}.
The list to the right of $\colonminus$ is the rule \define{body}. Bodies are defined by 
    recursion as follows, along with associated sets $\fvar$ (free variables).
    \medskip
    
    \noindent
    \begin{tabularx}{1.0\linewidth}{lLL>{\small}l}
  \textbf{Name}  & \textbf{Form}  & \fvar & \textbf{\normalsize Comment}\\\hline
  Ordinary atom & p(\TT, \vec{t}) & \var(\TT, \vec t) & $\TT$ time term, $p$ free predicate  \\
\pbox[t]{Comprehension \\[-0.5ex] \small with time term $x$} & p(x \circ \TT, \vec t) \STH B & \{x\} \cup \var(\TT, \vec t) & $\circ \in \{<, \le, >, \ge \}$, $B$ is a body\\
Built-in call & p(\vec t) & \var(\vec t) & $p$ is built-in predicate\\
Time comparison & s \circ t & \var(s,t) & $s$, $t$ time terms, $\circ \in \{<, \le, >, \ge \}$\\
Let special form & \LET(x, t) & \{x\}\cup\var(t)  \\
Choose special form & \CHOOSE(x, \mathit{ts}) & \{x\} \cup \var(\mathit{ts}) & $\mathit{ts}$ is a set of terms\\
Collect special form & \COLLECT(x, t \STH B) & \{x\} \\\hline
Positive body $\vec b$ & b_1,\ldots,b_k & \cup_{i=1..k} \fvar(b_i)  & $k \ge 0$, $b_i$ is one of above \\
Negative body literal & \NOT \vec b & \emptyset & $\vec b$ is non-empty positive body \\
Body $B$ & \hspace*{-2ex}\vec b,\NOT \vec{b}_1, \ldots,\NOT \vec{b}_n & \fvar(\vec{b}) & $n\ge 0$, and $\vec b$, $\vec b_j$  positive bodies\\\hline
\end{tabularx}
\medskip

A \define{positive body literal} is of one of the forms up to $\COLLECT$. Examples are below.
\begin{note}[Implicit quantification]
In a body $B$, the variables $\fvar(B)$ are implicitly existentially quantified in front
of that $B$.\footnote{The variables $\var(t)$ in the $\COLLECT$ special form have to be excluded
from that because they are quantified within their ``$\STH B$'' body
scope.  To avoid name
conflicts, we assume that $\var(t) \cap \fvar(B') = \emptyset $ for all bodies $B'$ such
that $B = B'$ or $ B$ occurs in $B'$.}
Rules may contain extra variables in negative body literals.
An example is the rule
$\fun{p}(\timevar, x) \colonminus \fun{q}(\timevar, x, y), \NOT (z < \timevar,\fun{r}(x,y,z) )$
which corresponds to the (universal quantification of the) formula
$ \fun{q}(\timevar, x, y) \land \neg \exists z . (z < \timevar \land \fun{r}(x, y, z)) \rightarrow \fun{p}(\timevar, x)$.
The extra variable $z$ will be picked up for existential quantification after
ground instantiating the rule body's $\fvar$s $\{\timevar, x, y \}$. If
$\gamma$ is such a grounding substitution then
indeed $\fvar((z <\timevar,\fun{r}(x, y, z))\gamma) = \{z \}$ as desired.
The formal definition of the possible model semantics below will make this precise.
\qed
\end{note}
A \define{normal rule} is an ordinary rule with one head literal ($m=1$ in \eqref{eq:rule}).
A \define{Horn} rule is a normal rule or a fail rule.
A \define{fact} is an ordinary rule with empty body ($k,n = 0$ in \eqref{eq:rule}) and is
simply written as $H$.  A rule $H \colonminus B$ is
\define{range-restricted} iff $\var(H) \subseteq \fvar(B)$.
A \define{(Fusemate) program} is a finite set of range-restricted 
and \emph{stratified} rules.

\paragraph{Stratification.}
The standard notion of stratification
(``by predicates'') means that the call graph of a program has no cycles going
through negative body literals~\cite{przymusinski89declarative}. 
Every strongly connected component of the call graph is called a stratum and contains the
predicates that are defined (in rule heads) mutually recursive with each other.  All
head predicates of the same rule are put into the same stratum.
Fusemate employs a weaker \define{stratification by time and by
  predicates (SBTP)}~\cite{Baumgartner:Fusemate:SystemDescription:CADE:2021}.
  With SBTP, every ordinary non-fact rule \eqref{eq:rule} must have an
ordinary body literal $b_i$, for some $1 \le i \le k$, with a \define{pivot}  variable
$\timevar$, such that every other time term in the head (body) is syntactically constrained
to $\ge$ ($\le$, respectively) than $\timevar$, and the literals within negative body literals
are syntactically constrained to be (a) 
$<$ than $\timevar$ or  (b) $\le$ than $\timevar$  and must be in a stratum strictly lower than the head stratum.
For example, the rule 
$\fun{p}(\timevar, x)  \colonminus \fun{q}(\timevar, x), \NOT ( \fun{r}(t, y), t\le\timevar)$
is SBTP if $\fun{r}$ is in a strictly lower stratum than $\fun{p}$, and 
$\fun{p}(\timevar, x)  \colonminus \fun{q}(\timevar, x), \NOT ( \fun{r}(t, y), t<\timevar)$
is SBTP even if $\fun{r}$ is in the same stratum as $\fun{p}$.
This has the effect that model computation can be done in
time/stratum layers in increasing (lexicographic) order using only already derived atoms.

Comprehension and $\COLLECT$ must be stratified for the same reason.
For the purpose of SBTP, a comprehension $p(x \circ \TT, \vec t) \STH B$ is taken
as if $p(x, \vec t)$ and $B$ were negative body literals, and
$\COLLECT(x, t \STH B)$ is taken as if $B$ were a negative body literal.

\subsection{Possible Models}
\label{sec:possible-models}
We need some preliminaries pertaining to the semantics of rules before formally
defining ``possible models''.
A \define{(rule) closure} is a pair
$(H \colonminus B,\beta)$ such that $\beta$ is a grounding substitution for $\fvar(B)$ called
\define{body matcher} in this context.
For a program $P$, its \define{full closure $\mathit{cl}(P)$} is  the set of all closures of all rules in $P$.

Full closures supplant the usual full ground instantiation of programs. They make it easy to
define rule semantics in presence of the special forms, comprehension operators, and
implicit existential quantification without full grounding. This works as follows.

An \define{interpretation} $I$ is a (possibly infinite) set of ordinary atoms.
Let  $I$ be an interpretation and $\beta$ a grounding substitution for some set of variables.
Let $B$ be a body as in~\eqref{eq:rule}. If $\fvar(B\beta) = \emptyset$ define
$I,\beta \models B$ iff $I,\beta \models b_1,\ldots,b_k$ and $I,\beta \models \NOT\vec{b}_j$ for all $j=1..n$,
where the following table provides the definitions for body literals:
    \medskip
    
    \noindent
    \begin{tabularx}{1.0\linewidth}{lLcL}
  \textbf{Name}  & \textbf{Form}  & & \textbf{Def} \\\hline
  Ordinary atom & I,\beta \models p(\TT, \vec{t}) & iff & p(\TT,\vec t)\beta \in I\\
      \pbox[t]{Comprehension \\[-0.5ex] \small with time term $x$} & I,\beta \models p(x < \TT, \vec t) \STH B & iff&
\pbox[t]{\small $x\beta$ is the maximal (latest) time point s.th.\\[-0.5ex] \small $x\beta < \TT
      \beta$, $I,\beta \models p(x,\vec t)$ and $I,\beta \models \exists B$. \\[-0.5ex] \small Accordingly for $\ge$, $<$, $\le$
      } \\
Built-in call & I,\beta \models p(\vec t)  & iff & \text{$p(\vec t)\beta$ evaluates to true} \\
Time comparison & I,\beta \models s \circ t & iff & s\beta \circ t\beta  \\
Let special form & I,\beta \models \LET(x, t) & iff & x\beta = t\beta \\
Choose special form & I,\beta \models \CHOOSE(x, \mathit{ts}) & iff & x\beta \in \mathit{ts}\beta \\
Collect special form & I,\beta \models \COLLECT(x, t \STH B) & iff & \pbox[t]{$x\beta  = \{ t\gamma \mid I,\beta\gamma \models B$  for some \\
  \hspace*{-2ex}grounding substitution $\gamma $ for $\fvar(B\beta)\}$}\\\hline
Positive body $\vec b$ & I,\beta \models b_1,\ldots,b_k & iff & I,\beta \models b_i\text{ for all } i = 1..k\\
Negative body literal & I,\beta \models  \NOT \vec b & iff & I,\beta \not\models \exists \vec b  \\\hline
\end{tabularx}
\medskip

In the table above, define \define{$I,\beta \models \exists B$} iff there is a grounding substitution $\gamma$ for $\fvar(B\beta)$ such that $I,\beta\gamma \models B$ ($\beta\gamma$ is
$\beta$ extended with bindings for the implicitly existentially quantified variables in$B\beta$).
For closures define $I \models (H \colonminus B, \beta)$
iff $I,\beta \not\models B$ or else $H$ is an ordinary head  $h_1 \vee \cdots h_m$ and $h_i\beta \in I$ for some $1
\le i \le m$. In this case we say that \define{$I$ satisfies $(H \colonminus B, \beta)$}.
An interpretation $I$ is a \define{model of a set $C$ of closures}, written as $I \models C$ iff $I$ satisfies every closure in $C$. 
It is \define{minimal} iff $J \not\models C$ for every $J \subsetneq I$. It is \define{supported}
iff for every $a \in I$ there is a $(h \colonminus B, \beta) \in C$ such that $a = h\beta$ and $I,\beta \models
B$. 

\begin{note}[Fixpoint iteration for DLPs~\cite{Sakama:PossibleModelSemanticsDisjunctiveDatabases:DOOD:89}]
  \label{note:fixpoint-iteration}
The possible model semantics
~\cite{Sakama:PossibleModelSemanticsDisjunctiveDatabases:DOOD:89,Sakama:Inoue:PossibleModels:JAR:94}
assigns to a disjunctive logic program sets of Horn programs and takes their
intended models as the possible models of the disjunctive program.
The Horn programs represent all possible ways of making one or
more head literals true, for every disjunctive rule.  As a propositional example, the 
disjunctive program $\{ \fun{a} \colonminus \fun{b},\ \fun{a} \vee \fun{c} \colonminus \fun{b},\ \fun{b}\colonminus\NOT \fun{d} \} $ is split into the
Horn programs
$\{ \fun{a} \colonminus \fun{b},\ \fun{b}\colonminus\NOT \fun{d} \} $ and
$\{ \fun{a} \colonminus \fun{b},\ \fun{c} \colonminus \fun{b},\ \fun{b} \colonminus \NOT \fun{d}\} $. The possible models are
$\{\fun{a}, \fun{b} \}$ and $\{\fun{a}, \fun{b}, \fun{c} \}$. Non-ground programs have to be fully ground-instantiated
using the program's (possibly infinite) Herbrand base first.

As explained in~\cite{Sakama:PossibleModelSemanticsDisjunctiveDatabases:DOOD:89},
the possible models of such ground-instantiated \define{stratified} programs can be constructed by
iterated fixpoint computation along the program's stratification. For each stratum, in
ascending order, the rules with a head predicate from that stratum are
evaluated in the model so far, up to that stratum, and, only if necessary, made true by adding the head to the
model, until fixpoint. In general this construction requires
transfinite induction with a limit ordinal at each stratum. \qed
\end{note}
From a practical (Fusemate) perspective we are mostly interested in finite fixpoints for
making model computation effective.
We start with a definition for the possible models splitting operator in terms of closures.
\begin{definition}[Split program closure]
  \label{def:split-program}
Let $P$ be a program and $\mathit{cl}(P)$ its full closure. A \define{split
  program closure of $P$} is obtained from $\mathit{cl}(P)$ by replacing every closure  $(h_1 \vee \cdots \vee h_m \leftarrow B,\beta)$ in $\mathit{cl}(P)$  by the
\define{split closures} $(h \leftarrow B,\beta)$, for every $h \in  S$, where $S$ is some non-empty subset of $\{h_1,\ldots,h_m\}$.
\end{definition}

\begin{note}[Flawed Definition of Possible Models]
\label{note:flaw}
  The original paper~\cite{Sakama:PossibleModelSemanticsDisjunctiveDatabases:DOOD:89}
  defines, in our words, an interpretation $I$ as a \define{possible model} of a program
  $P$ iff $I$ is a minimal supported model of some split program of $P$. Unfortunately,
  there is a flaw in this definition.  To explain, by way of example, take the program
  $P = \{ \fun{a} \colonminus \fun{a},\ \fun{b}\colonminus\NOT \fun{a} \} $.  It has two
  minimal supported models, $I_{\text{good}} = \{b\}$ and $I_{\text{bad}} = \{a\}$ which
  are exactly the possible models of $P$ according to this definition.
However, while $I_{\text{good}}$ will be computed by fixpoint iteration, $I_{\text{bad}}$ will be
not. Clearly,  $I_{\text{bad}}$ is not intended as a possible
model in~\cite{Sakama:PossibleModelSemanticsDisjunctiveDatabases:DOOD:89}. The example,
thus, disproves the completeness claim for fixpoint iteration
in~\cite{Sakama:PossibleModelSemanticsDisjunctiveDatabases:DOOD:89} (Theorem~3.1).

The flaw stems from requiring minimality of models \emph{as a whole}. A fixed definition
needs to match the iterated fixpoint construction, which computes minimal (and
supported) models on a \emph{per stratum} basis. In the example, only $\emptyset$ is a minimal model of the first stratum $\{ a
\colonminus a \}$  which is extended to the minimal model $\{ b \}$ of $P$. 
The \emph{perfect model semantics} of~\cite{przymusinski89declarative} achieves that and 
will be used below as a more suitable basis for defining possible models.
With that fix, a Theorem 3.1
in~\cite{Sakama:PossibleModelSemanticsDisjunctiveDatabases:DOOD:89} will hold.
\qed
\end{note}

\begin{definition}[Possible models, adapted
  from~\cite{Sakama:PossibleModelSemanticsDisjunctiveDatabases:DOOD:89} and corrected]
  \label{def:possible-model}
  An interpretation $I$ as a \define{possible model} of $P$ iff $I$ is a perfect
  model of some split program closure of $P$.
\end{definition}

\subsection{Fusemate Soundness and Completeness}
\label{sec:fusemate-soundness-completeness}
We wish to apply the fixpoint model construction (Note~\ref{note:fixpoint-iteration}) to Fusemate programs. For this to work,
rules must be \emph{monotonic} and \emph{compact}. 
\begin{definition}
  \label{def:rule-monotonicity-compactness}
  Let $(H \colonminus B, \beta)$ be an ordinary rule closure. It is \define{monotonic} iff
  for all $I$ and $J \supseteq I$ such that every atom in $J\setminus I$ is in the same stratum as $H\beta$,
  if $I,\beta \models B$ then $J,\beta \models B$. It is \define{compact} iff for all  $I$, if $I,\beta \models B$ then $J,\beta \models B$ for some finite $J \subseteq I$.
\end{definition}
In general, monotonicity of an operator guarantees the existence of a least fixpoint,
and compactness guarantees that it can be found by fixpoint iteration.
For satisfiable Horn programs, monotonicity entails the ``model intersection property'' which
entails the existence  of a unique minimal model.
These are all well-known standard results~\cite{Lloyd:87}, and the above definitions are
formulated in a way to make these results applicable. 

Fusemate rules are always monotonic. In particular for comprehension and
$\COLLECT$  this is due to stratification. However $\COLLECT$ is not always compact.
Given a body literal $\COLLECT(x, t \STH B)$, there could be
infinitely many substitutions $\gamma$ in the comprehension
$\{ t\gamma \mid I,\beta\gamma \models B \text{ for some grounding substitution $\gamma $ for $\fvar(B\beta)$} \}$.
Because infinite sets have no term representation, such a $\COLLECT$ literal renders its rule
body always unsatisfied, resulting in incompleteness. One possible way out is to make sure that the
variables in $t$ range only over finite domains, e.g., sets of constants.
With this fix, it follows that fixpoint
iteration (Note~\ref{note:fixpoint-iteration}) wrt.\ SBTP is sound and complete for possible
models of Fusemate programs  (Def.~\ref{def:possible-model}. The proof is an adaptation of
the corresponding one in~\cite{Sakama:PossibleModelSemanticsDisjunctiveDatabases:DOOD:89}.

Soundness and completeness of fixpoint iteration
holds in particular for \emph{finite} models. This suggests another ``fix'':
thanks to stratification, the mentioned incompleteness can occur only when
$I$ itself is infinite at a limit step in the fixpoint iteration.
Because computing (rather, finitely representing) infinite models is out of scope anyway,
it is safe to ignore the compactness problem for finite model computation.

\paragraph{Fusemate.}
Fusemate implements a bottom-up model computation procedure in the style of hyper
tableaux~\cite{Baumgartner:Furbach:Niemelae:HyperTableau:JELIA:96} in a stratified way
(SBTP).  
The Fusemate main loop computes body matchers $\beta$
of bodies $B$ of program rules $H \colonminus B$  against a current branch (a
model candidate) and closes it or branches out according to possible models splitting.
Each new branch is for a set $S$ in Def.~\ref{def:split-program} and receives all  $h\beta$ for $h \in S$.\footnote{Body matcher are
  represented internally in the Scala runtime system without explicit grounding.}
This constructs tableau in a depth-first left-to-right order.
Body matcher computation is made more practical by guaranteed left-to-right evaluation of
bodies.
This helps to avoid unexpected undefinedness of comprehensions and special forms.
For example, in the body of $\fun{r}(\TV, \mathit{xs}) \colonminus \fun{q}(\TV, y),  \COLLECT(\mathit{xs}, x \STH (\fun{p}(\TV,x), x > y))$
  the $\COLLECT$ special form binds the variable $\mathit{xs}$ to the list of all $x$ such that
  $\fun{p}(\TV,x)$ and $x > y$ hold, where $y$ has \emph{already} been bound by the preceding $\fun{q}(\TV,
  y)$. See~\cite{Baumgartner:Fusemate:SystemDescription:CADE:2021} for a formal definition
  of left-to-right body matcher computation.

  Other than that, Fusemate model computation follows the abstract fixpoint
  computation procedure (see Note~\ref{note:fixpoint-iteration}) for finite
  interpretations. This entails \emph{finite model soundness}: if Fusemate terminates on a program $P$ with an
  open exhausted branch then this branch contains a finite possible model of 
  $P$.
  It also entails \emph{finite model  completeness}: if every possible model of $P$ is finite then Fusemate will
  compute each of them in its open exhausted branches.
  A formal theorem for these results could be given but is not stated here because it
  would require more formalization.

  Fusemate's termination behavior could be improved with a breadth-first strategy, however
  at the expense of one-branch-at-a-time space efficiency. In the programs below this is
  not a problem.

  \section{Description Logic Interface}
\label{sec:dl-interface}
Fusemate can be used as a description logic (DL) reasoner by mapping a DL
knowledge base into a logic program and running that program for satisfiability~\cite{Baumgartner:Fusemate:SystemDescription:CADE:2021}.
This section makes that reasoner callable from rules, but other DL reasoners could be coupled, too.
It describes the syntax, semantics, and soundness and completeness properties  of the
coupling, and it discusses related work.

The DL terminology follows~\cite{Baader:etal:introduction-description-logics:2017}.  To
summarize, a DL knowledge base KB consists of a TBox and an ABox.  A TBox $T$ is a set of
GCIs (general concept inclusions), each of the form $C \sqsubseteq D$ where $C$ and $D$ are DL
\define{concept expressions}, or just \define{concepts}.  
An ABox $A$ is a set of \define{ABox assertions}, i.e., concept assertions and role assertions of the
forms $a \: C$ and $(a, b)\:r$, respectively, where $a$ and $b$ are individuals and $r$ is
a role. Fusemate
currently implements \ALCIF, which is \ALC extended with inverse roles and functional
roles. A \define{role}, hence, is either a role name $n$ or an inverse role name
$n^{-1}$. Roles can be declared as functional (right-unique). As usual,
KB-satisfiability is assumed to be decidable and concept formation must be closed under
negation, so that query entailment can be reduced to KB
unsatisfiability as follows. Given a KB $(A, T)$ and an ABox $Q$, the
\define{(ground) query}, define \define{$(A, T) \models_{\fun{DL}} Q$} iff the KB entails $Q$
wrt.\ the usual first-order logic semantics of description logics, or,
equivalently: for all $a \: C \in Q$ the KB $(A \cup \{ a \: \neg C \}, T)$ is unsatisfiable and for all
$(a,b) \: r \in Q$ the KB $(A \cup \{ a \: \forall\, r. \neg B,\ b \:  B \},T)$ is
unsatisfiable, where $B$ is a fresh concept name.

The coupling between the rules and the DL reasoner is \emph{two-way} and \emph{dynamic}:
it is two-way in the sense that rules can not only \emph{call} the DL reasoner wrt.\ a
fixed ABox and a TBox, the rules can also \emph{construct}  ABoxes during model
computation, individually in each possible model. It is \emph{dynamic} in the sense that ABox assertions are time-stamped, like
ordinary atoms, and also all earlier ABoxes are accessible by the rules.

\paragraph{Syntax.} Concepts and roles are treated as constants by the rule language while any free
ground term can be a DL  individual. More precisely, assume a DL signature whose
concept and role names are disjoint with the signature of the rule language. 
Let $t, t_1, t_2$ be free possibly non-ground terms, $C$ a concept, $r$ a role and $\TT$ a time term.
An \define{untimed DL-atom} is of the form $t \: C$ or $(t_1, t_2) \: r$. 
Let $\IsAAt/3$ and $\HasAAt/4$ be distinguished ordinary predicate symbols.
A \define{timed DL-atom} is an ordinary atom $\IsAAt(t, C, \TT)$ or
$\HasAAt(t_1, r, t_2, \TT)$, usually written as $t\: C \AT \TT$ or $(t_1, t_2)\: r \AT
\TT$, respectively.
Timed DL-atoms can appear in heads (and bodies) of ordinary rules. This allows to
create time-stamped ABoxes initially as sets of facts and dynamically during program execution.
For calling the DL reasoner, the rule language is extended by the following
\define{DL-call} special forms, where $T$ is a TBox, $A$ is an ABox, and $\vec{q}$ (``query'')
is a list of untimed DL-atoms.
\begin{xalignat*}{3}
T & \models \vec{q} &   \fun{DLISSAT}&(T) & \fun{DLISUNSAT}&(T) \\
(A, T) & \models \vec{q} &  \fun{DLISSAT}&(A, T) &  \fun{DLISUNSAT}&(A, T)
\end{xalignat*}
The free variables are $\fvar(\vec q)$ in the left column cases, otherwise empty.

\paragraph{Semantics.}
Logic programming considers syntactically different terms
as unequal. This is not enforced in DLs. Indeed, e.g.,
if $A = \{ (\fun{a},\fun{c})\: \fun{r},  (\fun{a},\fun{b})\: \fun{r}\}$ and $\fun{r}$
is a functional role then $A$ is satisfiable by making $\fun{b}$ and $\fun{c}$ 
\emph{equal}. To avoid such discrepancies, DL individuals are explicitly equipped with a
unique name assumption, as follows.

Given an ABox $A$,  let $K(A) = \{ a_1,\ldots a_n \}$ be the set of all
(``known'') individuals mentioned in
$A$ and define $\mathit{UNA}(A) = \{ a_i \: N_{(a_i,a_j)}, a_j \: \neg N_{(a_i,a_j)}  \mid a_i, a_j
\in K(A) \text{ and } 1 \le i < j \le n\}$. In that, $N_{(a_i,a_j)}$ are fresh concept
names. The set $\mathit{UNA}(A)$ specifies that all individuals in $A$  must be pairwise unequal ($\fun{a}$, $\fun{b}$ and $\fun{c}$ in
the example).

The definition of rule semantics in
Section~\ref{sec:stratified-logic-programs} is extended by DL-calls as follows:
$I, \beta \models ((A, T)  \models \vec{q})$ iff $(A \cup \mathit{UNA}(A) \cup \mathit{UNA}(\vec{q}\beta), T) \models_{\fun{DL}} \vec{q}\beta$
($\vec{q}\beta$ as a set);
$I, \beta \models \fun{DLISSAT}(A, T)$ iff $(A \cup \mathit{UNA}(A), T)$ is satisfiable;
$I, \beta \models \fun{DLISUNSAT}(A, T)$ iff $(A \cup \mathit{UNA}(A), T)$ is unsatisfiable.

For the DL-calls on the first line, let $\TV$ be the pivot variable of the rule containing the
DL-call and take $A = \mathit{abox}(I, \TV\,\beta)$ for the corresponding definition with explicit
$A$,  where 
$\mathit{abox}(I, d) = \{ t\: C \mid t\: C \AT d \in I \} \cup \{ (t_1,t_2)\: r \mid (t_1,t_2)\: r \AT d \in
I\}$ is the \define{induced ABox from $I$ at time $d$}. Intuitively, such a DL-call gets its ABox 
from the current interpretation by projection from its timed DL-atoms at the current time.

Notice the implicit dependency of an induced ABox on timed DL-atoms at pivot time. This is why for the
purpose of stratification every line one DL-call stands for the two subgoals
$\IsAAt(\_, \_, \TV)$ and $\HasAAt(\_, \_, \_, \TV)$.  For constant ABoxes on the second
line stratification is not an issue. (As such they are not very useful - but see
Example~\ref{ex:2} and the example in Section~\ref{sec:alltogether} below.)

With all that in place, the possibly model semantics for stratified programs defined in
Section~\ref{sec:possible-models} carries over to the DL coupling without change. Notice
that the semantics of the coupling is agnostic of the notion of (un)satisfiability and
entailment in the DL part. This way, the coupling respects the usual open world semantics
of DL reasoning. Notice also that it is possible that a program has a possible
model $I$ whose induced ABox is unsatisfiable with some TBox $T$. If this is not
desirable it is easy to reject such a model with a $\FAIL$ rule utilising a $\fun{DLISUNSAT}(T)$ call.

\paragraph{Soundness and completeness.}
Soundness and completeness carries over from
Section~\ref{sec:fusemate-soundness-completeness} with some caveats.
Incompleteness can arise due to potentially
infinite ABoxes induced at limit ordinals. With an interest in finite models
only, this issue can safely be ignored, as before. A more relevant issue
is monotonicity (Def.~\ref{def:rule-monotonicity-compactness}). $\fun{DLISSAT}$ calls
can be non-monotonic because first-order logic satisfiability is, of course, not always preserved when a KB
grows. This can lead to both incompleteness/unsoundness, depending on a positive/negative call context.
The other two forms are based on unsatisfiability, hence monotonic, and cause no problem.
With those only, iterated fixpoint computation and Fusemate model computation
are both sound and complete for finite possible models.

\paragraph{Related work.}
According to the classification in~\cite{Eiter2008}, ours  is a hybrid approach with a loose coupling between the
description logic and the rule reasoner. The coupling is done in a DL-safe
way~\cite{10.1007/978-3-540-30475-3_38}, in fact,
essentially, in a \emph{weakly} DL-safe~\cite{Rosati:etal:tight-integration:KR:2006} way
as in DL+Log.
DL+log~\cite{Rosati:etal:tight-integration:KR:2006} is among the most expressive languages
that combines rules with ontologies. DL+log rules can query a DL reasoner by taking
concept/role names as unary/binary predicates and using (in our terms) extra existentially
quantified variables in queries. With Fusemate rules one would equivalently use existential role
restrictions. Unlike DL+log, Fusemate allows DL-calls within default negation, cf.\ Example~\ref{ex:2}.
Most other hybrid languages, like the one in \cite{10.1007/978-3-540-30475-3_38} and
dl+Programs~\cite{EITER20081495} do not allow DL atoms in the head.
Unlike as in the other approaches, concepts and roles are \emph{terms} here and, hence, can
be quantified over in rules. This is advantageous for writing domain independent rules
involving concepts and roles, such as the event calculus in
Section~\ref{sec:event-calculus}.

\subsection{Example}
As a running example
we consider a highly simplified transport scenario. Boxes containing goods are loaded onto
a truck, moved to a destination, and unloaded again. The boxes can contain perishable
goods that require cooling, fruits, or non-perishable goods, toys. Boxes of the former
kind (and only those) can be equipped with temperature sensors and provide a temperature
value, which is classified as low (unproblematic) or high (problematic).  
A part of this domain is modelled in the description logic $\ALC$ extended with functional
roles. The following KB has a TBox on box properties (left), and an ABox on temperature classes (middle) and 
box properties (right):

{ \setlength{\jot}{1pt}
  \small
  \begin{xalignat*}{3}
  \mathsf{Box} & \sqsubseteq \forall\ \mathsf{Temp} . \mathsf{TempClass}  & 
  \mathsf{Low}  & : \mathsf{TempClass} & \mathsf{Box}_0 & : \mathsf{FruitBox}\\
  \mathsf{FruitBox} & \sqsubseteq \exists\ \mathsf{Temp} . \mathsf{TempClass} & \mathsf{High} & :
  \mathsf{TempClass} & \mathsf{Box}_1 & : \mathsf{FruitBox} \\
  \mathsf{ToyBox} & \sqsubseteq \neg \exists\ \mathsf{Temp} . \mathsf{TempClass} & & & \mathsf{Box}_2 & : \mathsf{Box}\\
  \mathsf{FruitBox} & \sqsubseteq \mathsf{Box} &  & & \mathsf{Box}_3 & : \mathsf{ToyBox}\\
  \mathsf{ToyBox} & \sqsubseteq \mathsf{Box} & & & \mathsf{Box}_4 & : \mathsf{Box} \sqcap \forall\ \mathsf{Temp} . \neg \mathsf{TempClass}\\
\rlap{\text{$\mathsf{Temp}$ is a functional role}} & &    & & \mathsf{Box}_5 & : \mathsf{Box} \sqcap \exists\ \mathsf{Temp} . \mathsf{TempClass}
\end{xalignat*}}

\begin{example}
\label{ex:1}
\noindent The ABox assertions can be represented as a program with facts timed at, say,
0 (``beginning of time''), e.g., $\fun{Box(5)} \: \mathsf{Box} \sqcap \exists\ \mathsf{Temp}
. \mathsf{TempClass} \AT 0$.\footnote{The concrete Fusemate syntax is
\LIN{IsAAt(Box(5), And(Box,Exists(Temp,TempClass)), 0)} but we stick with the better
readable ``:''-syntax. TBoxes have similar syntax and are typically bound to (Scala)
variables like $\fun{tbox}$ in the example.
In concrete syntax, free constant, function and predicate symbols start with a capital
letter, variables with a lower case letter. An underscore \LIN{_} is an anonymous
variable.}
Let $\fun{tbox}$ denote the TBox above. Some example rules with DL-calls are
\begin{lstlisting}
x : Box @ time $\colonminus$ x : _ @ time, tbox $\models$ x : Box
TempBox(time, box) $\colonminus$ box : Box @ time, tbox $\models$ box $\: (\exists\ \fun{Temp} . \fun{TempClass})$
KnownTempBox(time, box) $\colonminus$
    box : Box @ time, choose(temp, List(Low, High)), tbox $\models$ (box, temp) : Temp
\end{lstlisting}
The first rule materializes the $\fun{Box}$ concept. Any known individual at a given
$\fun{time}$ that is provable a $\fun{Box}$ will explicitly become a $\fun{Box}$ individual at
$\fun{time}$. While this is redundant for DL-reasoning, it comes in handy for rules.  For
example, the second rule applies to explicitly given $\fun{Box}$es at $\fun{time}$ that provably have a $\fun{Temp}$
attribute. Thanks to the first rule, \LIN{TempBox(0, Box($i$))} is derivable for $i \in
\{\fun{0},\fun{1},\fun{5}\}$. (Recall that the ABox in the DL-call is formed from the timed DL-atoms
at pivot $\fun{time}$.) The third rule is a variation of the second rule and tests if a box
has a \emph{concrete} $\fun{Temp}$ attribute $\fun{Low}$ or $\fun{High}$ instead of \emph{some}. 
\qed
\end{example}

\begin{example}
\label{ex:2}
This is an example for a stratified DL-call within default negation and explicit ABox:
\begin{lstlisting}
ColdBox(time, box) $\colonminus$
    box : Box @ time,
    not (t < time, (I.aboxAt(t), tbox) $\models$ box : Box, (box, High) : Temp)
\end{lstlisting}
According to this rule, a $\fun{box}$ is a $\fun{ColdBox}$ at a given $\fun{time}$ if it never
provably was a $\fun{Box}$ in the past with a $\fun{High}$ temperature.
The (Scala) expression \LIN{I.aboxAt(t)} can be used in Fusemate to retrieve the
  induced abox at time $\fun{t}$ from the current interpretation $I$.\footnote{Access to $I$ is
    unusual for logic programming systems.
    See~\cite{Baumgartner:Fusemate:SystemDescription:CADE:2021} for a discussion of this features.}
Notice that $t$ is strictly earlier than $\fun{time}$ which renders the DL-call stratified.

An example for the $\fun{DLISUNSAT}$ DL-call is in the rule
\LIN{fail $\colonminus$ Now(time), DLISUNSAT(tbox)}. This rule abandons a current model
candidate if its induced abox at the current $\fun{time}$ ``Now'' is inconsistent with $\fun{tbox}$.
\qed
\end{example}

\section{Event Calculus Embedding}
\label{sec:event-calculus}
The event calculus (EC) is a logical language for representing and reasoning about actions and
their effects~\cite{DBLP:journals/ngc/KowalskiS86,Shanahan1999}. At its core, effects are
fluents, i.e., statements whose truth value can change over time, and the
event calculus provides a framework for specifying the effects of actions in terms of
initiating or terminating fluents. 

Many versions of the EC exists, see \cite{Miller2002} for a start. The approach below makes do
with a basic version that 
is inspired by the discrete event calculus in~\cite{Mueller:2004a} with integer time.
The event calculus of~\cite{Mueller:2004a} is operationalized by
translation to propositional SAT. Its 
implementation in the ``decreasoner'' is geared for efficiency and can be used to solve
planning and diagnosis tasks, among others. The version below is 
tailored for the model computation tasks mentioned in the introduction, where a fixed sequence
of events at given timepoints can be supposed.\footnote{Actually, events can be inserted
  in retrospect using Fusemate's revision operator, restarting the model computation from
  there. The paper~\cite{Baumgartner:PossibleModelsSpringer:IJCAR:2020} already has a
  ``supply-chain'' example for that.} It rests on minimal model semantics and stratified
default negation. Most of it is not overly specific to Fusemate, and
answer set programming encodings of the event calculus, 
e.g. \cite{event-calculus-ASP:JAIR:2012}, should be applicable as well.


The rest of this section explains the EC/DL integration grouped into ``axiom sets'':
\begin{itemize}
\item Domain independent EC axioms: principles of actions initiating/terminating fluents
\item Domain independent EC/DL integration axioms: ABox assertions as fluents
\item Domain dependent axioms: initial situation and concrete actions effects
\item Concrete actions: events driving the model computation
\item Fusemate specific rules
\end{itemize}

\paragraph{Domain independent axioms.}
The EC main syntactic categories are \define{Fluents} and \define{Actions}, both given
via designated sub-signatures of the term signature. They are used with 
\define{EC-predicates} in intended sorting as follows:
\begin{xalignat*}{2}
\fun{Initiates} & : \Time \times \fun{Action} \times \fun{Fluent} &
\fun{Initiated} & : \Time \times \fun{Fluent} \\
\fun{Terminates} & : \Time \times \fun{Action} \times \fun{Fluent} &
\fun{Terminated} & : \Time \times \fun{Fluent} \\
\fun{StronglyTerminates} & : \Time \times \fun{Action} \times \fun{Fluent} &
\fun{StronglyTerminated} & : \Time \times \fun{Fluent} \\[1ex]
\fun{HoldsAt} & : \Time \times \fun{Fluent}  & \fun{Happens} & : \Time \times \fun{Action} 
\end{xalignat*}
The EC was originally introduced as a Prolog logic program.
The following \define{domain independent rules} are similar but modified for stratified bottom-up
model computation.
Some rules use a ``strong negation'' operator $\NEG$ which can be applied to
ordinary atoms in the body or the head. Fusemate implements the usual
semantic~\cite{gelfond91classical} which amounts to adding the rules
$\FAIL \colonminus p(\timevar, \vec{x}), \NEG p(\timevar, \vec{x})$ for every ordinary
predicate $p$.
\begin{lstlisting}
Initiated(time+1, f) $\colonminus$ Happens(time, a), Initiates(time, a, f) // H1
Terminated(time+1, f) $\colonminus$ Happens(time, a), Terminates(time, a, f) // H2
StronglyTerminated(time+1, f) $\colonminus$ Happens(time, a), StronglyTerminates(time, a, f) // H3
Terminated(time, f) $\colonminus$ StronglyTerminated(time, f) // H4

HoldsAt(time, f) $\colonminus$ Initiated(time, f), not Terminated(time, f) // EC3
neg(HoldsAt(time, f)) $\colonminus$ StronglyTerminated(time, f), not Initiated(time, f) // EC4

HoldsAt(time, f) $\colonminus$ Step(time, prev), HoldsAt(prev, f), not Terminated(time, f) // EC5
neg(HoldsAt(time, f)) $\colonminus$ Step(time, prev), neg(HoldsAt(prev, f)), not Initiated(time, f) // EC6
\end{lstlisting}
In the rules above, the variable \fun f stands
for fluents and \fun a for actions.
The axioms H1 -- H3 specify the dependencies between fluents and actions in general.
The distinction between \fun{Initiates} and \fun{Initiated} was made
for being able to distinguish between initiation by actions (``\emph{loading} a
box on a truck \emph{initiates} the box being on the truck'') and initiation as a matter
of circumstances or their consequences (``smoke \emph{initiated} alarm bell ringing'').

The core relation is \LIN{HoldsAt(time,f)} which can
hold true at \fun{time} because \fun f  is \fun{Initiated} at \fun{time} (EC3), or was true at
the previous time step but not \fun{terminated} (EC5, frame axiom). Similarly for the negated case.
Notice the difference between \fun{Terminated} and \fun{StronglyTerminated}. The former
removes \LIN{HoldsAt(time, f)} from the model, the latter inserts
\LIN{neg(HoldsAt(time,f))} into it. That is, this is a three-valued logic. With default negation 
one can distinguish the three cases.

Notice that fluents are initiated or terminated in H1 -- H3 with a delay of one time step. This was
done so that the \fun{Initiates} and \fun{Terminates} predicates can be defined in a
stratified way in terms of \fun{HoldsAt} at the current \fun{time}.
Without the delay SBTP would be violated in such cases.
The increase in time
will not cause non-termination of model computation because H1 -- H3  are conditioned on
events happening (as long as there are only finitely many events).

\subsection{Linking Description Logic with the Event Calculus}
Section~\ref{sec:dl-interface} introduced timed DL-atoms for specifying (timestamped)
ABoxes.  Typically, ABox assertions
should be preserved over time unless there is reason for change. Examples are 
the initial ABox assertions in Example~\ref{ex:1} and role assertions in
Example~\ref{ex:4} below.
This immediately suggests to utilize the event calculus for treating ABox assertions as
fluents. The following explains this in more detail.

\paragraph{Domain independent axioms.}
From now on, \emph{un}timed DL-atoms are allowed in fluent positions.
Untimed DL-atoms are enough because fluents occur within EC-predicate atoms which by
themselves provide the time. 
The following axioms are added as domain independent axioms to restore the timed DL-atom
versions of the fluents:
\begin{lstlisting}
x : c @ time  $\colonminus$ HoldsAt(time, x : c) // DL1
x : Neg(c) @ time  $\colonminus$ neg(HoldsAt(time, x : c)) // DL2
(x, y): r @ time $\colonminus$ HoldsAt(time,(x, y) : r) //DL3
\end{lstlisting}
Notice the use of variables $c$ and $r$ in concept and role positions, which makes it possible
to formulate these axioms independent of a concrete DL KB.  The DL2 axiom expresses strongly
negated concept membership equivalently by membership in the negated concept.

The axioms DL1 -- DL3 are obviously reasonable in any domain. 
Their converse is not, however. Not everything holding true at a point in time should by default
extend into the future, e.g., a person's birthday.

\paragraph{Domain dependent axioms.}
Domain dependent axioms comprise fluents that hold initially and specifications of
action effects in terms of initiation and termination of fluents.
An example for the former is the fact for $\fun{Box(5)}$  in Example~\ref{ex:1}, which could be
rewritten as
$\fun{HoldsAt}(0, \fun{Box(5)} \: \mathsf{Box} \sqcap \exists\ \mathsf{Temp}. \mathsf{TempClass})$.
\begin{example} 
  \label{ex:3}
  The following rules specify the effects of $\fun{Load}$ and $\fun{Unload}$ actions of boxes in
  terms of these boxes being $\fun{OnTruck}$.
\begin{lstlisting}
Initiates(time, Load(box), OnTruck(box)) $\colonminus$ box : Box @ time
StronglyTerminates(time, Unload, OnTruck(box)) $\colonminus$ HoldsAt(time, OnTruck(box))
\end{lstlisting}
  The first rule makes sure in its body that only boxes that exist at a time can be
  loaded. The second rule concludes that all boxes loaded will definitely not be not on
  the truck after unload. All other boxes are untouched.
  Notice that the  $\fun{OnTruck}$ fluent is not a DL concept (it doesn't have to be).
  \qed
\end{example}

\paragraph{Concrete actions.} What is still missing are concrete actions happening for
triggering the model computation in the combined Rules/DL/EC domain model. 
In the running example we consider the following scenario unfolding:
\medskip

{\small
\setlength{\jot}{1pt}
\noindent\begin{tabularx}{\linewidth}{r@{\qquad}XXXXX}
  \textbf{Time}  &
  10 & 20 & 30 & 40 & 50
  \\\hline
  \textbf{Action} &
\pbox[t]{Load $\fun{Box}_0$\\ Load $\fun{Box}_1$} & 
Load $\fun{Box}_2$& 
\pbox[t]{Load $\fun{Box}_3$\\ Load $\fun{Box}_4$} & 
& 
Unload 
  \\
  \textbf{Sensor} &
$\fun{Box}_0: -10\degree$ & 
$\fun{Box}_2 : 10\degree$ & 
$\fun{Box}_0 : 2\degree$ & 
$\fun{Box}_0 : 20\degree$ & 
\end{tabularx}
}

These actions are easily represented as facts, e.g., \LIN{Happens(10, Load(Box(0)))}.
The temperature measurement at time
  20 for $\fun{Box(2)}$ becomes \LIN{Happens(20, SensorEvent(Box(2), 10))}.

\paragraph{Concrete domains.}
Real-world applications require reasoning with concrete domains (numeric types, strings,
etc). Extending DLs with concrete domains while preserving satisfiability is possible only
under tight expressivity bounds. See~\cite{DBLP:conf/aiml/Lutz02} for a survey.
One way to mitigate this problem is to use rules and built-ins for concrete domains and
to pass symbolic abstractions to the DL reasoner. 
\begin{example}
  \label{ex:4}
  This rule demonstrates abstracting a concrete box temperature sensor reading as a $\fun{Temp}$
  attribute. 
    \begin{lstlisting}
Initiates(time, SensorEvent(box, temp), (box, High) : Temp) and
   Terminates(time, SensorEvent(box, temp), (box, Low) : Temp) ) $\colonminus$
    Happens(time, SensorEvent(box, temp)), temp > 0
    \end{lstlisting}
  The given action
\LIN|Happens(20, SensorEvent(Box(2), 10))|
with the rule above and rules H1 and H6 will derive
\LIN|HoldsAt(21, (Box(2),High) : Temp)|.
From that, with DL1 and the rules in Example~\ref{ex:1}, $\fun{Box(2)}$  will
  become a $\fun{TempBox}$ and even a $\fun{KnownTempBox}$ from time $\fun{21}$ onwards.
\qed
\end{example}

\paragraph{Fusemate specific rules.}
Fusemate provides the user with a number of non-standard operators,
see~\cite{Baumgartner:Fusemate:SystemDescription:CADE:2021}. One of them is the aggregation
operator $\fun{COLLECT}$.
\begin{example}
  \label{ex:5}
  Consider the rule
    \begin{lstlisting}
Unloaded(time+1, boxes) $\colonminus$
	Happens(time, Unload),
	COLLECT(boxes, box STH HoldsAt(time, OnTruck(box)))
    \end{lstlisting}
This rule aggregates all unloaded boxes into one set, $\fun{boxes}$, one tick after
$\fun{Unload}$ time. It is not formulated as a fluent to make it a time\emph{point}
property. In the example, the $\fun{Unload}$ happens at time 50, which leads
  to \LIN{Unloaded(51, Set(Box(0), Box(1), Box(2), Box(3), Box(4))}. Notice that these
  are exactly the boxes loaded over time, at timepoints 10, 20, and 30.\qed
\end{example}

\subsection{Ramification Problem}
\label{sec:ramification-problem}
The ramification problem is concerned with indirect consequences of an action. Such
consequences could be in conflict with facts holding at the time of the action or other consequences. 
This problem is particularly prominent
in the combination with DL, where effects (i.e., fluents) can be entailed implicitly by
the DL KB, and possibly
in an opaque way. Trying to terminate such a fluent can be futile, as it could be
re-instated implicitly or explicitly by materialization. 

A good example is the entailment  of \LIN{TempBox(0, Box(0))} as discussed in Example~\ref{ex:1}. 
Suppose we wish to re-purpose $\fun{Box(0)}$ and no longer use it for temperature sensitive
transport. In terms of the modelling, $\fun{Box(0)}$  shall no longer belong to the (entailed)
concept $\exists \fun{Temp} . \fun{TempClass}$.

The ramification problem has been extensively researched in the EC literature, see~\cite{Shanahan1999}. For instance, one could impose state constraints, if-and-only
if conditions, so that terminating an entailed fluent propagates down; or one could use
effect constraints that propagate termination of actions to other actions.
A first attempt in this direction is a rule that terminates a fluent that entails the
property to be removed:
\begin{lstlisting}
Terminated(time+1, (box, temp) : Temp)) $\colonminus$
	RemoveTemp(time, box),  // Some condition for removing box Temp
	(box, temp) : Temp @ time // Attribute to be removed
\end{lstlisting}
  This rule works as expected for $\fun{Box}_2$ after explicitly having received a
\fun{Temp}-attribute at time 20, cf.\ Example~\ref{ex:4}.
It does not work, however, for,
e.g., $\fun{Box}_0$. As a $\fun{FruitBox}$, $\fun{Box}_0$ has a $\fun{Temp}$ attribute implied by
the TBox. 

One way to fix this problem \emph{in the running example} is to terminate \emph{all} concept assertions for the $\fun{box}$
as any of them might entail a $\fun{Temp}$ attribute, and only retain that it is a $\fun{Box}$:
\begin{lstlisting}
(Terminated(time,  box : concept) and Initiated(time,  box : Box))$\colonminus$
	RemoveTemp(time, box),  // Some condition for removing box Temp
	box : concept @ time, concept != Box // Concept to be removed
// Similar rule for removing role assertions omitted
\end{lstlisting}
While this measure achieves the desired effect, it may also remove box properties that
could be retained, e.g., the size of the box (if it were part of the example, that is).

The KB revision problem has been studied extensively in database and AI 
settings. For DLs, there are algorithms for \emph{instance level updates} of an ABox,
where, in first-order logic terms, the ABox is a set of ground atoms over known
individuals, see~\cite{DBLP:journals/jair/GiacomoORS21}. Very recently, Baader
etal~\cite{Baader-et-al:compliant-optimisations:ISWC:2020,Baader-et-al:abox-repair:CADE:2021}
devised algorithms for semantically optimally revising ABoxes that may contain
quantifiers (e.g. $\fun{Box}_5$ in the  running example). 
All these result are for lightweight description logics, though.

\section{Putting it all Together}
\label{sec:alltogether}
This section completes the running example with rules for diagnostic reasoning.
Suppose a given subset of the boxes $\{\fun{Box}_0,\ldots,\fun{Box}_5\} $ is unloaded at the
destination.  We are interested in determining the status of the delivery and computing
possible models as explanations under these constraints:
\begin{enumerate}
\item If there is no unloaded box with known high temperature then the status is OK.
\item If some unloaded box has a known high temperature then this box has been 
  tampered with or the truck cooling is broken.
  \item If some unloaded box has a known low temperature then the truck cooling is not broken
    (because a broken cooling would affect all boxes).
  \item If all unloaded boxes with a temperature sensor can consistently be assumed to have
    high temperature then box tampering can be excluded (because broken cooling is the
    more likely explanation).
\end{enumerate}
The following rules determine the status of the delivery as ``ok'' or ``anomalous''. There are two
cases of anomalies, (a) the truck cooling is broken or (b) some box has been tampered
with. The rules feature disjunctive heads,
strong negation, DL-calls, Scala builtin calls and the set datatype. 
\begin{lstlisting}
OK(time) $\colonminus$ Unloaded(time, boxes), not Anomaly(time, _)

Anomaly(time, TamperedBox(box)) or Anomaly(time, BrokenCooling) $\colonminus$
    Unloaded(time, boxes),
    (box, High) : Temp @ time,
    boxes $\ni$ box

neg(Anomaly(time, BrokenCooling)) and neg(Anomaly(time, TamperedBox(box))) $\colonminus$
    Unloaded(time, boxes),
    (box, Low) : Temp @ time,
    boxes $\ni$ box

fail $\colonminus\ \ $  Anomaly(time, TamperedBox(box)),
    Unloaded(time, unloadedBoxes),
    COLLECT(boxes, box STH (TempBox(time, box), unloadedBoxes $\ni$ box)), 
    LET(assertions, boxes map { (_, High) : Temp }), // unloaded boxes ascribed High Temp 
    DLISSAT(I.aboxAt(time) ++ assertions, tbox)
\end{lstlisting}
  The first rule makes the delivery ok in absence of any anomaly.  The second rule
  observes an anomaly if some unloaded box has a \fun{High} temperature. The anomaly could
  be either type, or both, this rule makes a guess.  The third and the fourth rule are
  eliminating guesses.  The third rule says that the truck cooling is not broken if 
 evidenced by the existence of a \fun{Low} temperature box. Moreover, each of these boxes
 has not been tampered with.
  The fourth rule is the most
  interesting one. It eliminates a tampered-box anomaly by considering all unloaded boxes
  that are known to be equipped with temperature sensors.  The rationale is that if
  \emph{all} these boxes can consistently be assumed to have \fun{High} temperature then
  box tampering is unlikely (broken cooling is more likely).

  This reasoning is achieved by collecting in line 15 in the \fun{boxes} variable the
  mentioned boxes (\fun{TempBox} was defined is Example~\ref{ex:1}). Line 16 assigns to a
  variable \fun{assertions} the value of the stated Scala expression for constructing \fun{High}
  temperature role assertions for \fun{boxes}. Finally, the DL-call on line 17 checks the
  satisfiability of the KB consisting of the current abox temporarily extended with
  \fun{assertions} and the static TBox.
  It is important to know that $\fun{fail}$ rules
  are always tried last for a fixed current $\fun{time}$, after all
  ordinary rules.  This way, the usages of \fun{COLLECT} and \fun{DLISSAT} in
  the last rule \emph{are} stratified.
  
The correct diagnosis is \LIN|Anomaly(51, BrokenCooling)|.
In the course of events, the \fun{TempBoxes} are $\fun{Box}_0$, $\fun{Box}_1$,
$\fun{Box}_2$, and $\fun{Box}_5$ ($\fun{Box}_2$ becomes one only at time 20.) The unloaded
boxes at time 50  are $\fun{Box}_0$, $\fun{Box}_1$, $\fun{Box}_2$, and $\fun{Box}_4$.  In their
intersection, $\fun{Box}_0$ and $\fun{Box}_2$ have $\fun{High}$  $\fun{Temp}$ values, which gives rise
to an anomaly. Only the box $\fun{Box}_1$ has an unknown $\fun{Temp}$  value, which is
consistent with $\fun{High}$ and, hence, excludes a $\fun{TamperedBox}$ anomaly.
Moreover, for every box, neither a $\fun{TamperedBox}$  anomaly nor a negated $\fun{TamperedBox}$
anomaly is derived.

If the $\fun{Box}_0$ sensor reading at time 40 is changed from 10 to -10 then the
diagnosis is
\begin{lstlisting}
Anomaly(51, TamperedBox(Box(2)))
neg(Anomaly(51, TamperedBox(Box(0))))
neg(Anomaly(51, BrokenCooling))
\end{lstlisting}
Both diagnosis are the only possible models in each case
and nothing is known about $\fun{Box}_1$. The Fusemate runtime is approx.\ 4 seconds in
each case on a modern PC. The main bottleneck is lack of performance of the coupled
DL-reasoner, which is a proof-of-concept implementation only.

\section{Conclusions}
This paper introduced a knowledge representation language that, for the first time, combines the event
calculus with description logic in a logic programming framework for model computation.
The paper demonstrated the interplay of these three components by means of an elaborated
example.

Results are in parts at an abstract level. They include conditions for finite-model
soundness and completeness of the rules/DL reasoner coupling that are re-usable in other
systems that support stratification in a similar way
(\cite{locallyStratifiedProgramsGreedy:Zaniolo:2015}, e.g.).

The diagnosis rules in
Section~\ref{sec:alltogether}, among others, utilized Fusemate's specific set
comprehension operator (\fun{COLLECT}) and might be hard to emulate in other systems. It
might be possible to run the example in this paper with an expressive system like
DLV~\cite{DLV:ACM:2006} without too many changes.

The modelling in the example emphasised the possibility to distinguish between absent,
unknown or known attribute values, which was enabled by the description
logics/rules integration. 
One might want to go a step further and add ``dynamic existentials'' to the
picture. These are unknown or implicit actions that must have existed to cause observed
effects. 
Recovered or speculating such actions can be expressed
already with the (implemented) belief revision framework
of~\cite{Baumgartner:PossibleModelsSpringer:IJCAR:2020}.
Experimenting with that within the framework here is future work.

The perhaps most pressing open issue is the EC ramification problem
(Section~\ref{sec:ramification-problem}),
which is particularly pronounced with the DL integration into the EC.
Recent advances on ABox updates might help~\cite{Baader-et-al:compliant-optimisations:ISWC:2020,Baader-et-al:abox-repair:CADE:2021}.

\paragraph{Acknowledgements.} I am grateful to the reviewers for their constructive feedback.

\bibliography{final}

\end{document}